\documentclass[%
 aip,
 amsmath,amssymb,
 reprint,%
]{revtex4-1}

\usepackage{graphicx}
\usepackage{dcolumn}
\usepackage{bm}
\usepackage{enumerate}

\usepackage[utf8]{inputenc}
\usepackage[T1]{fontenc}
\usepackage{mathptmx}
\usepackage{etoolbox}
\usepackage{enumitem}

\DeclareMathAlphabet{\mathcal}{OMS}{cmsy}{m}{n}

\usepackage{xcolor}
\newcommand{\dd}{\textup{d}}
\def\eps{\varepsilon}

\def\P{\mathbb{P}}
\def\R{\mathbb{R}}

\newcommand{\Markov}[2]{\underset{#1}{\overset{#2}{\rightleftharpoons}}}

\newcommand{\Db}{D_{\mathrm{b}}}

\newcommand{\Dc}{D}
\newcommand{\gb}{\gamma_{\mathrm{b}}}
\newcommand{\X}{\mathbf{X}}

\newcommand{\x}{\mathbf{x}}

\makeatletter
\def\@email#1#2{%
 \endgroup
 \patchcmd{\titleblock@produce}
  {\frontmatter@RRAPformat}
  {\frontmatter@RRAPformat{\produce@RRAP{*#1\href{mailto:#2}{#2}}}\frontmatter@RRAPformat}
  {}{}
}%
\makeatother
\begin{document}

\preprint{AIP/123-QED}

\title[Diffusive flux into a stochastically gated tube]{Diffusive flux into a stochastically gated tube}
\author{Sean D. Lawley}%
 \email{lawley@math.utah.edu.}
\affiliation{ 
Department of Mathematics, University of Utah, Salt Lake City, UT 84112 USA 
}%

\date{\today}

\begin{abstract}
Diffusion-influenced reactions in the presence of gates which randomly open and close have been studied for decades in a variety of biophysical and biochemical scenarios. The diffusive flux from a large bulk reservoir to the end of a narrow tube with a stochastically gated entrance has been previously estimated. In this paper, we extend this gated flux estimate to be valid if (i) the tube is not necessarily narrow and/or (ii) the diffusivity differs in the tube versus the bulk. Extension (i) is challenging because it entails a nontrivial three-dimensional geometry. Extension (ii) is challenging because it introduces multiplicative noise. We derive an explicit flux estimate formula and prove that it is exact in certain parameter regimes. We further use stochastic simulations to show that the estimate remains accurate across a very broad range of parameters. Our results differ from prior work on extensions (i) and (ii). 
\end{abstract}

\maketitle

\section{\label{sec:introduction}Introduction}

Many biological processes combine diffusion with a stochastic gate that randomly opens and closes. For example, the binding of ligands to proteins often requires that the molecules diffuse near each other and that the protein be in an ``open'' state which permits binding\cite{zhou2010rate}. Another example is the membrane transport of charged particles via voltage-gated or ligand-gated ion channels \cite{hille2001}.   
A further example is insect respiration, which depends on the diffusion of oxygen and carbon dioxide through tracheal tubes that connect to the ambient environment via holes in the exoskeleton that rapidly open and close during the flutter phase of the discontinuous gas exchange cycle\cite{wigglesworth1965principles}.  
In light of these and other biological systems, theorists have sought to quantify the rate of such ``stochastically gated'' diffusion processes in a wide variety of geometries and biophysical scenarios\cite{mccammon1981gated, szabo1982, zhou1996theory, spouge1996single, PB2, PB3, PB4, gopich2016reversible, lawley2019electrodiffusive, lawley2019dtmfpt, mercado2021first}. 

\begin{figure}[t]
\centering
\includegraphics[width=1\linewidth]{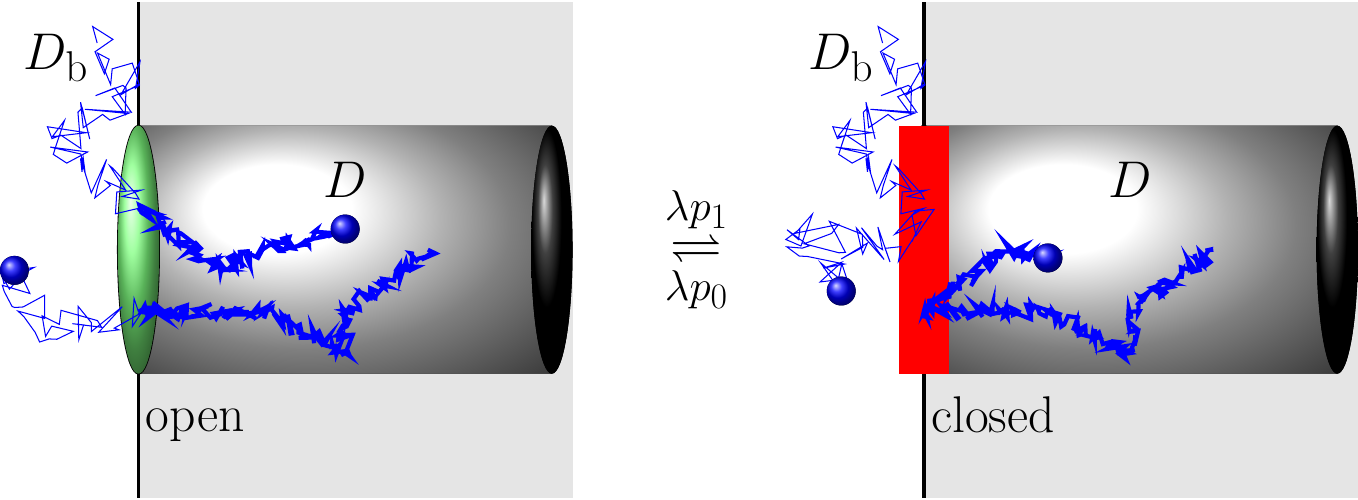}
\caption{Particles can diffuse from a bulk reservoir into a tube and get absorbed at the right end of the tube or wander back into the bulk. The tube is stochastically gated, meaning that its entrance randomly switches between being open (left panel) and closed (right panel). Particles can freely enter and exit the tube at the left end when the gate is open, whereas particles can neither enter nor exit at the left end when the gate is closed. Particles diffuse with diffusivity $\Db$ in the bulk and $\Dc$ in the tube.}
\label{fig:schem}
\end{figure}

The diffusive flux into a stochastically gated tube was studied in a student thesis\cite{lawleythesis} and published in Ref.~\citenum{lawley15sima}. To briefly explain, consider a cylindrical tube of radius $a$ and length $L$. Suppose diffusing particles enter the left end of the tube ($x=0$) from a large bulk reservoir and are absorbed at the right end of the tube ($x=L$). See Figure~\ref{fig:schem} for an illustration.  
Refs.~\citenum{lawleythesis, lawley15sima} assumed that the entrance to the tube at $x=0$ stochastically switches between open and closed states according to a Markov process,
\begin{align*}
    \text{open}\Markov{\lambda p_0}{\lambda p_1}\text{closed},
\end{align*}
where $\lambda p_0$ and $\lambda p_1$ denote the respective opening and closing rates and $p_0=1-p_1$ denotes the fraction of time that the tube is open. Assuming that the tube is narrow (i.e.\ $a/L\ll1$) and that particles have a constant diffusivity $D$, Refs.~\citenum{lawleythesis} and \citenum{lawley15sima} found that the average flux of particles absorbed at $x=L$ is
\begin{align}\label{eq:thesis}
    J
    &:=\bigg(1+\frac{p_1}{p_0}\frac{\tanh(\gamma)}{\gamma}\bigg)^{-1}\,J^{\text{open}},   
\end{align}
where $\gamma=\sqrt{\lambda L^2/D}$ compares the switching rate to the timescale of diffusion in the tube, and $J^{\text{open}}$ is the steady-state flux if the gate is always open. 

The formula in \eqref{eq:thesis} predicts that the gated flux is always greater than the naive estimate of simply multiplying the ``always open'' flux $J^{\text{open}}$ by the fraction of time that the gate is open (i.e.\ $J>p_0 J^{\text{open}}$). 
In fact, if the switching is fast compared to the diffusion timescale (i.e.\ $\gamma\gg1$), then \eqref{eq:thesis} predicts that the gated flux approaches the flux in the case that the gate is always open,
\begin{align}\label{eq:fast0}
    \lim_{\gamma\to\infty}J
    =J^{\text{open}}.
\end{align}
The prediction in \eqref{eq:fast0} is counterintuitive because it means that even if the gate is open only a small fraction of time ($p_0\ll1$), the flux can be nearly as large as if the gate is always open. That is, switching fast is the same as always being open. 
It was argued that this helps explain the flutter phase in insect respiration\cite{lawleythesis, lawley15sima, lawley2020spiracular, lawley2022water}. 

A very interesting study\cite{berezhkovskii2016diffusive} sought to extend the flux formula in \eqref{eq:thesis} to be valid if
\begin{enumerate}[label=(\roman*), start=1]
    \item the tube is not necessarily narrow (i.e.\ $L\not\gg a$),
    \item the diffusivity differs in the tube versus the bulk.
\end{enumerate}
Extension (i) is challenging because it means that the particle concentration is governed by three-dimensional partial differential equations (i.e.\ the one-dimensional approximation used to derive \eqref{eq:thesis} is no longer valid). Extension (ii) is challenging because it introduces multiplicative noise to the diffusion process, which entails subtleties in the stochastic calculus\cite{mannella2012ito}. 

In this paper, we provide an alternative analysis to make extensions (i) and (ii). We derive an explicit flux estimate and prove that it is exact in certain parameter regimes. We further use stochastic simulations to show that our flux estimate remains accurate across a very broad range of parameters. Our flux estimate differs from the flux estimate derived in Ref.~\citenum{berezhkovskii2016diffusive}. 

The rest of the paper is organized as follows. We begin in section~\ref{sec:firstpedagogical} with a pedagogical model that illustrates some concepts of stochastic gating and diffusion in an exactly solvable setting. We present and analyze our model of diffusion into a gated tube in section~\ref{sec:tube}. Section~\ref{sec:simulations} compares our flux estimate to stochastic simulations. Section~\ref{sec:predictions} explores some predictions of our flux estimate.  We conclude in section~\ref{sec:discussion} by comparing our flux estimate with prior analyses\cite{lawleythesis, lawley15sima, berezhkovskii2016diffusive}. An appendix collects some more technical aspects of our analysis.

\section{\label{sec:firstpedagogical}A pedagogical model}

We start with a pedagogical model of stochastically gated diffusion in a simple one-dimensional setting which is exactly solvable. 

\subsection{Always open}

Consider a particle diffusing in the one-dimensional interval $[-{{a}},L]$ with $a>0$ and $L>0$. Suppose the diffusivity $D(x)$ depends on the particle location $x$. In particular, suppose the diffusivity is a constant $\Db>0$ on the left side of the interval and a (possibly different) constant $\Dc>0$ on the right side,
\begin{align}\label{eq:Dx}
    D(x)
    =\begin{cases}
        \Db & \text{if }x<0,\\
        \Dc & \text{if }x>0.
    \end{cases}
\end{align}
To anticipate the model in section~\ref{sec:tube}, we interpret the $[-{{a}},0)$ as the ``bulk'' and $(0,L]$ as the ``tube.''

As emphasized by the It\^{o} versus Stratonovich controversy\cite{van1981ito, mannella2012ito, sokolov2010ito, vaccario2015, tung2025escape, tung2026stochastic}, a space-dependent diffusivity requires one to specify the interpretation of the multiplicative noise. For a parameter $\alpha\in[0,1]$, suppose that the probability density $p(x,t)$ that the particle is at location $x\in[-{{a}},L]$ at time $t$ satisfies the following continuity conditions at $x=0$,
\begin{align}\label{eq:cont0}
\begin{split}
    \Db^{1-\alpha}p(0^-,t)
    &=\Dc^{1-\alpha}p(0^+,t),\\
    \Db\partial_xp(0^-,t)
    &=\Dc\partial_xp(0^+,t),
\end{split}    
\end{align}
where $g(0^\pm)=\lim_{x\to0^\pm}g(x)$ denotes one-sided limits. In \eqref{eq:cont0}, the parameter $\alpha\in[0,1]$ specifies the interpretation of the multiplicative noise\cite{bressloff2017temporal}. The most common\cite{mannella2012ito} interpretations are $\alpha=0$ and $\alpha=1/2$, which correspond respectively to the It\^{o}\cite{ito1944stochastic} and Stratonovich\cite{stratonovich1966new} conventions. A third convention is $\alpha=1$, which is sometimes called the kinetic, isothermal, or H{\"a}nggi-Klimontovich convention\cite{hanggi1982stochastic}.

Let $P(x)$ be the probability that the particle hits $L>0$ before $-{{a}}<0$ given that the particle is initially at $x\in[-{{a}},L]$. Hence, $P(L)=1$, $P(-{{a}})=0$, and\cite{bressloff2017temporal}
\begin{align*}
    P''
    &=0,\quad x\in(-{{a}},0)\cup(0,L).
\end{align*}
Furthermore, $P$ satisfies the continuity conditions at $x=0$,
\begin{align}\label{eq:contPsimple}
    P(0^-)
    =P(0^+),\quad
    \Db^\alpha P'(0^-)
    =\Dc^\alpha P'(0^+).
\end{align}
It is straightforward to solve this problem and obtain
\begin{align}\label{eq:Pngsimple}
    P^{\text{open}}(0)
    =\frac{1}{1+\rho},
\end{align}
where $\rho$ compares the lengthscales of the ``tube'' $(0,L]$ and the ``bulk'' $[-a,0)$ to the change in diffusivity,
\begin{align}\label{eq:rhodef1d}
    \rho
    =(L/{{a}})(\Db/\Dc)^\alpha.
\end{align}
The superscript ``open'' in \eqref{eq:Pngsimple} emphasizes that \eqref{eq:Pngsimple} is for the case that the gate is always open.

Notice that \eqref{eq:Pngsimple} implies that if $\alpha>0$ (i.e.\ any non-It\^{o} convention), then increasing the ratio $\Db/\Dc$ decreases the probability that the particle hits the right end of the interval before the left end. This result stems from the so-called\cite{serov2020statistical} ``spurious'' force introduced by any non-It\^{o} space-dependent diffusivity which pushes particles toward regions of higher diffusivity.

\subsection{Gating}

Suppose now that there is a stochastic gate at $x=0$. In particular, suppose that the gate opens and closes according to a Markov jump process,
\begin{align}\label{eq:markovsimple}
    \text{open}\quad 0\Markov{\lambda_1}{\lambda_0}1\quad\text{closed},
\end{align}
where the respective opening and closing rates are
\begin{align}
\begin{split}\label{eq:lambda01}
    \lambda_1
    &=p_0\lambda,\quad\text{(opening rate)}\\
    \lambda_0
    &=p_1\lambda,\quad\text{(closing rate)}
\end{split}    
\end{align}
and $p_0$ and $p_1=1-p_0$ are the respective stationary probabilities of finding the gate open and closed,
\begin{align}
\begin{split}\label{eq:stationary}
    p_0
    &=\frac{\lambda_1}{\lambda_0+\lambda_1}
    =\frac{\lambda_1}{\lambda}\quad\text{(probability open)},\\
    p_1
    &=\frac{\lambda_0}{\lambda_0+\lambda_1}
    =\frac{\lambda_0}{\lambda}\quad\text{(probability closed)},
\end{split}    
\end{align}
and $\lambda=\lambda_0+\lambda_1$ parameterizes the overall ``switching rate'' between being open and being closed.

Let $P_0(x)$ (respectively, $P_1(x)$) be the probability that a particle that starts at $x\in[-{{a}},L]$ will reach $L$ before $-{{a}}$, given that the gate is initially open (respectively, closed). These probabilities satisfy the following backward Kolmogorov (or backward Fokker-Planck) equations for $x\in(-{{a}},0)\cup(0,L)$,
\begin{align}\label{eq:backward}
\begin{split}
    0
    &=D(x)\Delta P_0-\lambda_0(P_0-P_1),\\
    0
    &=D(x)\Delta P_1-\lambda_1(P_1-P_0),
\end{split}    
\end{align}
where $D(x)$ is in \eqref{eq:Dx}, with boundary conditions,
\begin{align*}
    P_0(-{{a}})=P_1(-{{a}})=0,\quad P_0(L)=P_1(L)=1.
\end{align*}
At $x=0$, $P_0$ satisfies the continuity conditions in \eqref{eq:contPsimple},
\begin{align}\label{eq:P0cont}
    P_0(0^-)
    =P_0(0^+),\quad
    \Db^\alpha P_0'(0^-)
    =\Dc^\alpha P_0'(0^+),    
\end{align}and $P_1$ satisfies no flux conditions,
\begin{align}\label{eq:P1noflux}
    P_1'(0^-)=0=P_1'(0^+).
\end{align}
Diagonalize \eqref{eq:backward} by introducing
\begin{align}
    P
    &=p_0P_0+p_1P_1,\label{eq:Pdefn}\\
    Q
    &=P_0-P_1.\label{eq:Qdefn}
\end{align}
The functions $P$ and $Q$ satisfy the following decoupled differential equations for $x\in(-{{a}},0)\cup(0,L)$,
\begin{align}
    \Delta P
    &=0,\label{eq:Pbackward}\\
    D(x)\Delta Q
    &=\lambda Q.\label{eq:Qbackward}
\end{align}
Furthermore, $P$ and $Q$ satisfy boundary conditions,
\begin{align*}
    P(-{{a}})=Q(-{{a}})=0,\quad P(L)=1,\quad Q(L)=0.
\end{align*}
The functions $P$ and $Q$ are decoupled, except for their continuity conditions at $x=0$,
\begin{align*}
    P(0^-)+p_1Q(0^-)
    &=P(0^+)+p_1Q(0^+),\\
    \Db^\alpha(P'(0^-)+p_1Q'(0^-))
    &=\Dc^\alpha(P'(0^+)+p_1Q'(0^+)),\\
    P'(0^-)-p_0Q'(0^-)
    &=0=P'(0^+)-p_0Q'(0^+),
\end{align*}
which follow from \eqref{eq:P0cont}-\eqref{eq:P1noflux} and the following relations which are equivalent to \eqref{eq:Pdefn}-\eqref{eq:Qdefn},
\begin{align*}
    P_0
    &=P+p_1Q,\\
    P_1
    &=P-p_0Q.
\end{align*}
It is straightforward to solve this problem and obtain
\begin{align}\begin{split}\label{eq:Pgatedsimple}
    P(0^-)
    &=\frac{1}{1+\frac{p_1}{p_0}\tanh({\gb})/{\gb}}P_0(0)\\
    &=\Big[1+\frac{p_1}{p_0}\frac{\tanh ({\gb})}{{\gb}}+\rho \Big(1+\frac{p_1}{p_0}\frac{\tanh (\gamma)}{\gamma}\Big)\Big]^{-1},
\end{split}    
\end{align}
where $\rho$ is in \eqref{eq:rhodef1d}
and $\gamma$ and ${\gb}$ compare the switching rate to the diffusion timescales in the ``tube'' and the ``bulk,''
\begin{align}
    \gamma
    &=\sqrt{\lambda L^2/\Dc},\label{eq:gamma1d}\\
    {\gb}
    &=\sqrt{\lambda a^2/\Db}.\label{eq:gammab1d}
\end{align}

\subsection{\label{sec:opposite1D}Takeaways from the pedagogical model}

We now summarize some key takeaways from this simple, exactly solvable model. Importantly, these takeaways persist in the more complicated model in section~\ref{sec:tube}.

First, in addition to the open probability $p_0$, the three important dimensionless parameters are $\rho$ in \eqref{eq:rhodef1d} and $\gamma$ and ${\gb}$ in \eqref{eq:gamma1d}-\eqref{eq:gammab1d}.

Second, the slow switching limit of the splitting probability in \eqref{eq:Pgatedsimple} is simply $p_0$ multiplied by the splitting probability in the case that the gate is always open,
\begin{align}\label{eq:slowsimple}
    \lim_{\gamma,{\gb}\to0}P(0)
    =p_0P^{\text{open}}(0),
\end{align}
where $P^{\text{open}}(0)$ is in \eqref{eq:Pngsimple}. 

Third, in the fast switching limit, the splitting probability in \eqref{eq:Pgatedsimple} is identical to the splitting probability in the case that the gate is always open,
\begin{align}\label{eq:fastsimple}
    \lim_{\gamma,{\gb}\to\infty}P(0)
    =P^{\text{open}}(0).
\end{align}

Fourth, the splitting probability in \eqref{eq:Pgatedsimple} depends strongly on the multiplicative noise parameter $\alpha\in[0,1]$ if $\Db\neq\Dc$.

\begin{figure}[t]
\centering
\includegraphics[width=1\linewidth]{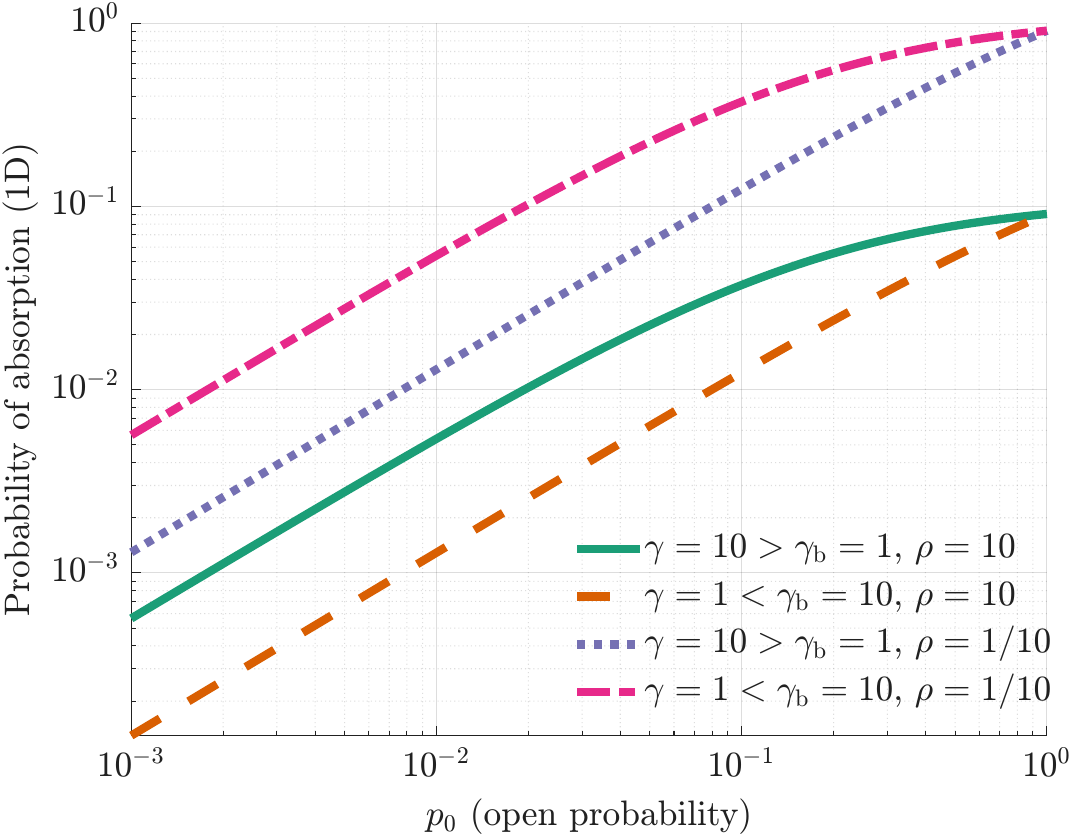}
\caption{Plot of $P(0^-)$ in \eqref{eq:Pgatedsimple} as a function of the open probability $p_0$ in different parameter regimes. See section~\ref{sec:opposite1D} for details.}
\label{fig:fastslow1D}
\end{figure}

Fifth, the different diffusivities ($\Db$ and $\Dc$) and the different lengthscales ($a$ and $L$) mean that the effective switching rate can be very different on opposite sides of the gate (i.e.\ it is possible to have $\gamma\gg\gb$ or $\gamma\ll\gb$). This fact can produce interesting and non-intuitive results. This is illustrated in Figure~\ref{fig:fastslow1D}, which plots $P(0^-)$ in \eqref{eq:Pgatedsimple} as a function of the open probability $p_0$. We emphasize two non-intuitive features of Figure~\ref{fig:fastslow1D}. First, the probability of absorption at $x=L$ is larger if $\gamma\gg\gb$ compared to $\gamma\ll\gb$ if $\rho\gg1$ (green and orange curves),
\begin{align}\label{eq:largerho1D}
    P(0^-)\big|_{\gamma=10>\gb=1}
    >P(0^-)\big|_{\gamma=1<\gb=10}\quad\text{if }\rho=10.
\end{align}
Second, the opposite is true if $\rho\ll1$ (purple and pink curves),
\begin{align}\label{eq:smallrho1D}
    P(0^-)\big|_{\gamma=10>\gb=1}
    <P(0^-)\big|_{\gamma=1<\gb=10}\quad\text{if }\rho=1/10.
\end{align}

To rationalize the result in \eqref{eq:largerho1D}, suppose $\rho=10$. In this case, it is unlikely that the particle will reach $x=L$ since $\rho=10\gg1$. The best chance of reaching $x=L$ is afforded by the gate closing once the particle reaches $0^+$, which is more likely if $\gamma$ is large. 

To rationalize the result in \eqref{eq:smallrho1D}, suppose $\rho=1/10$. In this case, any particle that reaches $0^+$ will likely reach $x=L$ since $\rho=1/10\ll1$. Therefore, $P_0(0)\approx1$, and thus $P(0^-)$ can be increased by increasing $P_1(0^-)$, which is accomplished by increasing $\gb$.

Finally, we emphasize that the four dimensionless parameters ($p_0$, $\rho$, $\gamma$, and ${\gb}$) can be varied independently through various regimes of $\alpha\in[0,1]$ and the six dimensional parameters $a$, $L$, $\Db$, $D$, $\lambda_0$, and $\lambda_1$. For example, if $a/L=\eps\ll1$ and $\Db/\Dc=\eps^4\ll1$, then
\begin{align}
    \gb/\gamma
    &=\eps^{-1}\gg1,\quad \rho=\eps^{-1}\gg1\quad\text{if }\alpha=0,\label{eq:regime1}\\
    \gb/\gamma
    &=\eps^{-1}\gg1,\quad \rho=\eps^{3}\ll1\quad\text{if }\alpha=1.\label{eq:regime2}
\end{align}
As another example, if $a/L=\eps^{-1}\gg1$ and $\Db/\Dc=\eps^{-4}\gg1$, then
\begin{align}
    \gb/\gamma
    &=\eps\ll1,\quad \rho=\eps\ll1\quad\text{if }\alpha=0,\label{eq:regime3}\\
    \gb/\gamma
    &=\eps\ll1,\quad \rho=\eps^{-3}\gg1\quad\text{if }\alpha=1.\label{eq:regime4}
\end{align}

\section{\label{sec:tube}Tube model}

We now formulate and analyze a three-dimensional model. The ``tube'' has length $L>0$, diameter $2a$, and cross-sectional shape $a\Gamma\subset\R^2$, where $\Gamma\subset\R^2$ is a dimensionless shape with dimensionless diameter $\text{diam}(\Gamma)=2$ (recall that the diameter of a set is the maximum possible distance between any two points in that set). That is, the tube is defined to be
\begin{align*}
    \Omega_{\text{tube}}
    =\{(x,y,z)\in\R^3: x\in(0,L),\,(y,z)\in a\Gamma\}.
\end{align*}
We are most interested in the case that the tube is a cylinder of radius $a$ (in which case $\Gamma$ is the unit disk), but we proceed in terms of a general cross-sectional shape.

The ``bulk'' is all of three-dimensional space to the left of the tube,
\begin{align*}
    \Omega_{\text{b}}
    =\{(x,y,z)\in\R^3: x<0\}.
\end{align*}
Suppose that the right end of the tube at $x=L$ is absorbing. 
Furthermore, suppose that particles have diffusivities $\Dc$ and $\Db$ in the tube and the bulk, respectively.  
That is, the diffusivity is the piecewise constant function of $x$ in \eqref{eq:Dx}. Suppose that the change in diffusivity is interpreted with multiplicative noise parameter $\alpha\in[0,1]$ as in section~\ref{sec:firstpedagogical} (with $\alpha=0$, $\alpha=1/2$, and $\alpha=1$ corresponding to the It\^o, Stratonovich, and kinetic conventions, respectively).

The steady-state flux of diffusing particles to the absorbing right end of the tube at $x=L$ is the product of (a) the flux to the tube entrance at $x=0$ and (b) the probability $\mathcal{P}$ that a particle which reaches the tube entrance from far away in the bulk will eventually be absorbed at the right end of the tube. The flux to the tube entrance is
\begin{align*}
    2\pi C_0(\Gamma)a\Db c_\infty,
\end{align*}
where $c_\infty$ is the fixed particle concentration far away in the bulk reservoir and $C_0(\Gamma)$ is the electrostatic capacitance of the shape $\Gamma$ (capacitance is defined so that the capacitance of a unit disk is $2/\pi$). The geometric prefactor $2\pi  C_0(\Gamma)a$ can be derived via a simple  divergence theorem argument\cite{lawley2026escape}. Note that $2\pi C_0(\Gamma)a=4a$ for a cylindrical tube\cite{berg1977}. 
Hence, the steady-state flux to the right end of the tube is
\begin{align}\label{eq:JP}
    J
    =(2\pi C_0(\Gamma)a\Db c_\infty)\mathcal{P}.
\end{align}
Therefore, calculating the flux $J$ amounts to calculating the probability $\mathcal{P}$. 
We first consider the case that the tube entrance is always open in section~\ref{sec:nogating}. We then suppose that the tube entrance is stochastically gated in section~\ref{sec:gating}. 

\subsection{\label{sec:nogating}Always open}

Let $P^{\text{open}}(\x)=P(\x)$ be the probability that a particle that starts at $\x=(x,y,z)$ will eventually reach the right end of the tube (assuming the entrance to the tube is always open). The function $P$ is harmonic in the bulk and in the tube,
\begin{align}\label{eq:Pharmonic}
    \Delta P
    &=0,\quad \x\in\Omega_{\text{b}}\cup\Omega_{\text{tube}}.
\end{align}
Further, $P$ vanishes at far-field and is equal to unity at the right end of the tube,
\begin{align*}
    \lim_{\|\x\|\to\infty,\,x<0}P(x,y,z)
    &=0,\\
    P
    &=1,\quad x=L,\,(y,z)\in a\Gamma,
\end{align*}
satisfies continuity conditions at the tube entrance,
\begin{align}
\begin{split}\label{eq:contentrance}
    P(0^-,y,z)
    &=P(0^+,y,z)\quad (y,z)\in a\Gamma,\\
    \Db^\alpha\partial_xP(0^-,y,z)
    &=\Dc^\alpha\partial_xP(0^+,y,z)\quad (y,z)\in a\Gamma,
\end{split}    
\end{align}
and satisfies reflecting boundary conditions on the rest of the boundary of $\Omega_{\text{b}}\cup\Omega_{\text{tube}}$.

For a particle which reaches the tube entrance from far away in the bulk, the probability of absorption is
\begin{align}\label{eq:mathcalP}
    \mathcal{P}
    &:=\int_{a\Gamma} P(0,y,z)\, \mu(y,z)\,\dd y\,\dd z,
\end{align}
where $\mu(y,z)$ is the probability density of where the particle hits the tube entrance if it started far from the tube and is conditioned to hit the tube. That is, $\mu(y,z)$ is the so-called harmonic measure density\cite{grebenkov2015analytical}. If the tube is a cylinder of radius $a$, then $\mu$ has a density given by the following explicit function\cite{sneddon1966mixed} of the radius $r=\sqrt{y^2+z^2}\in[0,a)$,
\begin{align}\label{eq:hittingdensity}
    \mu(r)&=\frac{r}{2\pi a^2\sqrt{1-(r/a)^2}},\quad r^2=y^2+z^2<a^2.
\end{align}

For $x\in[0,L]$, define the uniform average of $P$ in the tube cross-section,
\begin{align}
    \overline{P}(x)
    &:=\frac{1}{|a\Gamma|}\int_{a\Gamma}P(x,y,z)\,\dd y\,\dd z,
\end{align}
where $|a\Gamma|=a^2|\Gamma|$ denotes the area of $a\Gamma\subset\R^2$. 
Hence, integrating \eqref{eq:Pharmonic}, using the two-dimensional divergence theorem and the reflecting boundary conditions on the boundary of $a\Gamma$, and solving the resulting ordinary differential equations yields
\begin{align*}
    \overline{P}(x)
    &=\overline{P}(0)+(x/L)(1-\overline{P}(0)).
\end{align*}
Taking the derivative yields
\begin{align}\label{eq:Pderiv}
    \overline{P}'(0^+)
    =(1-\overline{P}(0))/L>0.
\end{align}
Integrating \eqref{eq:Pharmonic} over a large hemisphere $H_R$ of radius $R\gg a$ and using the divergence theorem yields
\begin{align}\label{eq:inthemi}
    0
    =\int_{H_R}\Delta P\,\dd \x
    =\int_{\partial H_R^+}\partial_{\mathbf{n}}P\,\dd S
    +\int_{a\Gamma}\partial_x P(0^-,y,z)\,\dd y\,\dd z,
\end{align}
where $\partial H_R^+$ denotes the curved part of the boundary of the hemisphere $H_R$ and $\partial_{\mathbf{n}}$ denotes the normal derivative. 
The strong Markov property implies that $P$ has the following monopole decay at far-field\cite{lawley2026escape},
\begin{align*}
    P(\x)
    \sim \frac{C_0(a\Gamma)\mathcal{P}}{\|\x\|}\quad\text{as }\|\x\|\to\infty,\,x<0,
\end{align*}
and therefore
\begin{align*}
    \int_{\partial H_R^+}\partial_{\mathbf{n}}P\,\dd S
    \to
    -2\pi aC_0(\Gamma)\mathcal{P}<0
    \quad\text{as }R\to\infty.
\end{align*}
Furthermore, the second continuity condition in \eqref{eq:contentrance} and the derivative of $\overline{P}$ in \eqref{eq:Pderiv} imply
\begin{align*}
    \int_{a\Gamma}\partial_x P(0^-,y,z)\,\dd y\,\dd z
    &=\Big(\frac{\Dc}{\Db}\Big)^\alpha\int_{a\Gamma}\partial_x P(0^+,y,z)\,\dd y\,\dd z\\
    &=\Big(\frac{\Dc}{\Db}\Big)^\alpha|\Gamma|a\frac{a}{L}(1-\overline{P}(0)).
\end{align*}
Therefore,
\begin{align}\label{eq:exact}
    \mathcal{P}
    =\frac{1}{G(\Gamma)}\frac{(1-\overline{P}(0))}{\rho},
\end{align}
where $G(\Gamma)$ is the following dimensionless geometric factor,
\begin{align}\label{eq:G}
    G(\Gamma)
    =\frac{2\pi C_0(\Gamma)}{|\Gamma|}>0,
\end{align}
and $\rho$ is the following dimensionless comparison of the tube aspect ratio and the change of diffusivity at the tube entrance,
\begin{align}\label{eq:rhodeftube}
    \rho
    :=(L/a)(\Db/\Dc)^\alpha>0.
\end{align}
Note that the geometric factor in \eqref{eq:G} is $G(\Gamma)=4/\pi$ if the tube is a cylinder (i.e.\ if $\Gamma$ is the unit disk). 

While the relation \eqref{eq:exact} is exact, we do not have an explicit formula for $\overline{P}(0)$. To obtain an explicit approximation for $\mathcal{P}$, we suppose
\begin{align}\label{eq:keyapprox}
    \mathcal{P}
    \approx\overline{P}(0)
\end{align}
and replace $\overline{P}(0)$ by $\mathcal{P}$ in \eqref{eq:exact} and solve for $\mathcal{P}=\mathcal{P}^{\text{open}}$ to obtain
\begin{align}\begin{split}\label{eq:Papproxold}
    \mathcal{P}^{\text{open}}
    \approx\mathcal{P}_{\text{approx}}^{\text{open}}
    :=&\frac{1}{1+G(\Gamma)\rho}.
\end{split}    
\end{align}
The superscript ``open'' emphasizes that \eqref{eq:Papproxold} is for the case that the gate is always open. 
The only approximation used to obtain \eqref{eq:Papproxold} is \eqref{eq:keyapprox}. 

We make four comments about the approximation in \eqref{eq:Papproxold}. First, recalling the definitions of $\mathcal{P}$ and $\overline{P}(0)$, the approximation \eqref{eq:keyapprox} (and therefore the approximation \eqref{eq:Papproxold}) is accurate if the fate of a particle which arrives at the tube entrance from far away in the bulk ($\mathcal{P}$) is similar to the fate of a particle which starts at the tube entrance and uniformly in the tube cross-section ($\overline{P}(0)$). Indeed, rearranging the exact relation in \eqref{eq:exact} implies that the relative error in the approximation in \eqref{eq:Papproxold} is given exactly by
\begin{align*}
    \frac{\mathcal{P}^{\text{open}}-\mathcal{P}_{\text{approx}}^{\text{open}}}{\mathcal{P}_{\text{approx}}^{\text{open}}}
    &=\mathcal{P}^{\text{open}}-\overline{P}(0)\\
    &=\int_{a\Gamma}P(0,y,z)\Big(\mu(y,z)-\frac{1}{|a\Gamma|}\Big)\,\dd y\,\dd z.
\end{align*}
Hence, the approximation in \eqref{eq:Papproxold} is guaranteed to be accurate if $P(0,y,z)$ is nearly constant across the tube cross-section. 

Second, the approximation \eqref{eq:Papproxold} is exact in the $\rho\to\infty$ limit\cite{richardson2026yet},
\begin{align*}
    \lim_{\rho\to\infty}\mathcal{P}^{\text{open}}/\mathcal{P}_{\text{approx}}^{\text{open}}=1.
\end{align*}
Third, \eqref{eq:Papproxold} is exact in the $\rho\to0$ limit\cite{richardson2026yet},
\begin{align*}
    \lim_{\rho\to0}\mathcal{P}^{\text{open}}=\lim_{\rho\to0}\mathcal{P}_{\text{approx}}^{\text{open}}
    =1.
\end{align*}
Fourth, stochastic simulations have shown that for the case of a cylindrical tube\cite{richardson2026yet},
\begin{align*}
    \mathcal{P}^{\text{open}}\approx\mathcal{P}_{\text{approx}}^{\text{open}}\quad\text{for all $\rho>0$}.
\end{align*}

\subsection{\label{sec:gating}Gating}

Suppose that there is a stochastic gate at the tube entrance at $x=0$ that opens and closes according to the Markov process in \eqref{eq:markovsimple} with parameters $p_0$, $p_1=1-p_0$, $\lambda_0$, $\lambda_1$, and $\lambda=\lambda_0+\lambda_1$ defined in \eqref{eq:lambda01}-\eqref{eq:stationary}. 
Let $P_0(\x)$ (respectively, $P_1(\x)$) be the probability that a particle that starts at position $\x=(x,y,z)$ will reach the right end of the tube at $x=L$, given that the gate is initially open (respectively, closed). These probabilities satisfy the backward equations in \eqref{eq:backward} for $\x=(x,y,z)\in\Omega_{\text{b}}\cup\Omega_{\text{tube}}$, where $D(x,y,z)=D(x)$ is in \eqref{eq:Dx}, with far-field conditions,
\begin{align*}
    \lim_{\|\x\|\to\infty,\,x<0}P_0(\x)=\lim_{\|\x\|\to\infty,\,x<0}P_1(\x)=0,
\end{align*}
and boundary conditions at the right end of the tube,
\begin{align*}
    P_0(\x)
    =P_1(\x)
    =1,\quad x=L,\,(y,z)\in a\Gamma.
\end{align*}
At the entrance to the tube, $P_0$ satisfies the continuity conditions in \eqref{eq:contentrance} and $P_1$ satisfies no flux conditions,
\begin{align}\label{eq:noflux}
    \partial_x P_1(0^-,y,z)
    =\partial_x P_1(0^+,y,z)
    =0,\quad (y,z)\in a\Gamma.
\end{align}

If a particle arrives from the bulk at the entrance to the tube, then the gate will be in its stationary distribution ($p_0$ and $p_1$ in \eqref{eq:stationary}), and thus
\begin{align*}
    \mathcal{P}
    =p_0 \mathcal{P}_0
    +p_1 \mathcal{P}_1,
\end{align*}
where $\mathcal{P}_0$ and $\mathcal{P}_1$ are defined analogously to \eqref{eq:mathcalP}, 
\begin{align}
    \mathcal{P}_0
    &=\int_{a\Gamma} P_0(0,y,z)\,\mu(y,z)\,\dd y\,\dd z
    ,\nonumber\\
    \mathcal{P}_1
    &=\int_{a\Gamma} P_1(0^-,y,z)\,\mu(y,z)\,\dd y\,\dd z.
    \label{eq:mathcalP0P1}
\end{align}
We emphasize that $P_1$ is evaluated at $x=0^-$ in \eqref{eq:mathcalP0P1}, which in general differs from $P_1$ at $x=0^+$.

Define
\begin{align}
    \overline{P}_0(x)
    &:=\frac{1}{|a\Gamma|}\int_{a\Gamma}P_0(x,y,z)\,\dd y\,\dd z,\\
    \overline{P}_1(x)
    &:=\frac{1}{|a\Gamma|}\int_{a\Gamma}P_1(x,y,z)\,\dd y\,\dd z.
\end{align}
Hence, integrating \eqref{eq:backward}, using the two-dimensional divergence theorem and the reflecting boundary conditions on the boundary of $a\Gamma$, we find that $\overline{P}_0(x)$ and $\overline{P}_1(x)$ satisfy the backward equations in \eqref{eq:backward} 
with boundary conditions
\begin{align}\label{eq:bc}
    \overline{P}_1'(0^+)
    &=0,\quad
    \overline{P}_0(L)
    =\overline{P}_1(L)
    =1.
\end{align}
It is straightforward to solve \eqref{eq:backward} subject to \eqref{eq:bc} to obtain $\overline{P}_0(x)$ and $\overline{P}_1(x)$ as explicit functions of $\overline{P}_0(0)$. For our purposes, we need only $\overline{P}_0'(0)$, which turns out to be
\begin{align}\label{eq:P0deriv}
    \overline{P}_0'(0)
    =\frac{1}{L}\frac{1}{p_0}\frac{(1-\overline{P}_0(0))}{(1+\frac{p_1}{p_0}\frac{\tanh (\gamma)}{\gamma})}.
\end{align}

Thus, if we define $P=p_0P_0+p_1P_1$ and $Q=P_0-P_1$ as in \eqref{eq:Pdefn}-\eqref{eq:Qdefn}, then $P$ and $Q$ satisfy \eqref{eq:Pbackward}-\eqref{eq:Qbackward} for $\x\in\Omega_{\text{b}}\cup\Omega_{\text{tube}}$.
Integrating \eqref{eq:Pbackward} over a large hemisphere of radius $R\gg a$ and using the divergence theorem as in \eqref{eq:inthemi} yields
\begin{align*}
    0
    =\int_{H_R}\Delta P\,\dd \x
    =\int_{\partial H_R^+}\partial_{\mathbf{n}}P\,\dd S
    +\int_{a\Gamma}\partial_x P(0^-,y,z)\,\dd y\,\dd z.
\end{align*}
The strong Markov property implies that $P$ has the following monopole decay at far-field\cite{lawley2026escape},
\begin{align*}
    P(\x)
    \sim \frac{C_0(a\Gamma)\mathcal{P}}{\|\x\|}\quad\text{as }\|\x\|\to\infty,\,x<0,
\end{align*}
and therefore
\begin{align*}
    \int_{\partial H_R^+}\partial_{\mathbf{n}}P\,\dd S
    \to-2a\pi C_0(\Gamma)\mathcal{P}
    \quad\text{as }R\to\infty.
\end{align*}
Furthermore, since $P_0$ satisfies the second continuity condition in \eqref{eq:contentrance}, $P_1$ satisfies the no flux condition in \eqref{eq:noflux}, and $P=p_0P_0+p_1P_1$, we have that 
\begin{align*}
    \int_{a\Gamma}\partial_x P(0^-,y,z)\,\dd y\,\dd z
    &=p_0\int_{a\Gamma}\partial_x P_0(0^-,y,z)\,\dd y\,\dd z\\
    &=p_0\Big(\frac{\Dc}{\Db}\Big)^\alpha\int_{a\Gamma}\partial_x P_0(0^+,y,z)\,\dd y\,\dd z\\
    &=\Big(\frac{\Dc}{\Db}\Big)^\alpha|\Gamma|a\frac{a}{L}\frac{(1-\overline{P}_0(0))}{(1+\frac{p_1}{p_0}\frac{\tanh (\gamma)}{\gamma})},
\end{align*}
where we have used \eqref{eq:P0deriv} in the final equality. 
Therefore,
\begin{align}\label{eq:exactswitch}
    \mathcal{P}
    =\frac{1}{G(\Gamma)}\frac{1}{\rho}p_0 L \overline{P}_0'(0^+)
    &=\frac{1}{G(\Gamma)}\frac{1}{\rho}\frac{(1-\overline{P}_0(0))}{1+\frac{p_1}{p_0}\frac{\tanh (\gamma)}{\gamma}}.
\end{align}
Similarly, integrating equation \eqref{eq:Qbackward} for $Q=P_0-P_1$ and using the divergence theorem yields
\begin{align}
    0
    &=\lambda/\Db\int_{x<0} Q\,\dd \x
    -\int_{a\Gamma}\partial_x Q(0^-,y,z)\,\dd y\,\dd z\nonumber\\
    &=\lambda/\Db\int_{x<0} Q\,\dd \x
    -\int_{a\Gamma}\partial_x P_0(0^-,y,z)\,\dd y\,\dd z\nonumber\\
    &=\lambda/\Db\int_{x<0} Q\,\dd \x
    -\frac{2a\pi C_0(\Gamma)\mathcal{P}}{p_0},\label{eq:Qexact}
\end{align}
where $\int_{x<0} Q\,\dd \x$ denotes the volume integral of $Q(\x)=Q(x,y,z)$ over the entire bulk $\Omega_{\text{b}}$.

We emphasize that the relations  \eqref{eq:exactswitch} and \eqref{eq:Qexact} are exact. To obtain an explicit approximation to $\mathcal{P}$ from \eqref{eq:exactswitch}-\eqref{eq:Qexact}, we make two approximations. First, we approximate
\begin{align}\label{eq:keyusual}
    \mathcal{P}_0
    \approx\overline{P}_0(0),\quad
    \mathcal{P}_1
    \approx\overline{P}_1(0^-),
\end{align}
which is analogous to \eqref{eq:keyapprox}. Second, we use the approximation (derived in Appendix~\ref{sec:derivationofkey}),
\begin{align}\label{eq:Qintapprox}
    \int_{x<0} Q\,\dd \x
    &\approx\frac{2\pi a C_0(\Gamma)+{\gb}a|\Gamma|}{\lambda/\Db}\big(\overline{P}_0(0)-\overline{P}_1(0^-)\big),
\end{align}
which then yields the following approximation from \eqref{eq:Qexact},
\begin{align}\label{eq:anotherapprox}
    \big(2\pi a C_0(\Gamma)+{\gb}a|\Gamma|\big)\big(\overline{P}_0(0)-\overline{P}_1(0^-)\big)
    &\approx\frac{2a\pi C_0(\Gamma)\mathcal{P}}{p_0}.
\end{align}
Rearranging \eqref{eq:anotherapprox} and using that $\mathcal{P}=p_0\mathcal{P}_0+p_1\mathcal{P}_1$ and the approximation \eqref{eq:keyusual} yields
\begin{align}\label{eq:key}
    \mathcal{P}_1
    &\approx\frac{p_0\gamma_\text{b}}{G(\Gamma)+p_0\gamma_\text{b}}\mathcal{P}_0,
\end{align}
which is equivalent to
\begin{align*}
    \mathcal{P}
    &\approx \frac{1}{1+\frac{p_1}{p_0}\frac{G(\Gamma)}{G(\Gamma)+{\gb}}}\mathcal{P}_0
\end{align*}
Putting this together, we replace $\overline{P}_0(0)$ in \eqref{eq:exactswitch} by $\mathcal{P}_0$ and use \eqref{eq:key} to obtain the following explicit approximation of $\mathcal{P}$,
\begin{align}\label{eq:Pmain}
    \mathcal{P}
    \approx
    \mathcal{P}_{\text{approx}}
    &:=\bigg[\frac{G(\Gamma)+p_0{\gb}}{G(\Gamma)p_0+p_0{\gb}}+G(\Gamma)\rho \Big(1+\frac{p_1}{p_0}\frac{\tanh (\gamma)}{\gamma}\Big)\bigg]^{-1}.
\end{align}

We show in Appendix~\ref{sec:largerho} that the approximation \eqref{eq:Pmain} is exact as $\rho\to\infty$. In particular, we show that
\begin{align}\label{eq:P0vanish}
    \lim_{\rho\to\infty}\overline{P}_0(0)
    =0,
\end{align}
and therefore the exact relation \eqref{eq:exactswitch} gives the following exact asymptotic,
\begin{align}\label{eq:exactasymptotic}
    \mathcal{P}
    &\sim\mathcal{P}_{\text{approx}}
    \sim\frac{1}{G(\Gamma)}\frac{1}{(1+\frac{p_1}{p_0}\frac{\tanh (\gamma)}{\gamma})
    }\frac{1}{\rho}\quad\text{as }\rho\to\infty,
\end{align}
where $f\sim g$ denotes $f/g\to1$. 
Furthermore, the simulations in section~\ref{sec:simulations} show that \eqref{eq:Pmain} is accurate even if $\rho\not\gg1$ if the tube is a cylinder. We explore the implications of \eqref{eq:Pmain} in section~\ref{sec:predictions}.

\section{\label{sec:simulations}Stochastic simulations}

We now compare the theory of section~\ref{sec:tube} to stochastic simulations for a cylindrical tube (so that $G(\Gamma)=4/\pi$). By \eqref{eq:JP}, calculating the steady-state particle flux $J$ amounts to calculating the splitting probability $\mathcal{P}$. The simulation algorithms are described in Appendix~\ref{sec:algorithm}. 

We showed that the approximation $\mathcal{P}_{\text{approx}}$ of $\mathcal{P}$ in \eqref{eq:Pmain} is exact if $\rho\gg1$. We thus start in the opposite regime of $\rho\ll1$, in which case $\mathcal{P}_{\text{approx}}$ in \eqref{eq:Pmain} becomes
\begin{align}\label{eq:Prho0}
    \lim_{\rho\to0}\mathcal{P}_{\text{approx}}
    =\frac{p_04/\pi+p_0\gb}{4/\pi+p_0\gb}
    =p_0+\frac{p_1p_0{\gb}}{4/\pi+p_0{\gb}}.
\end{align}
In this $\rho\ll1$ case, any particle which hits the entrance to the tube at $x=0$ when the gate is open will necessarily be absorbed. 

\begin{figure}[t]
\centering
\includegraphics[width=1\linewidth]{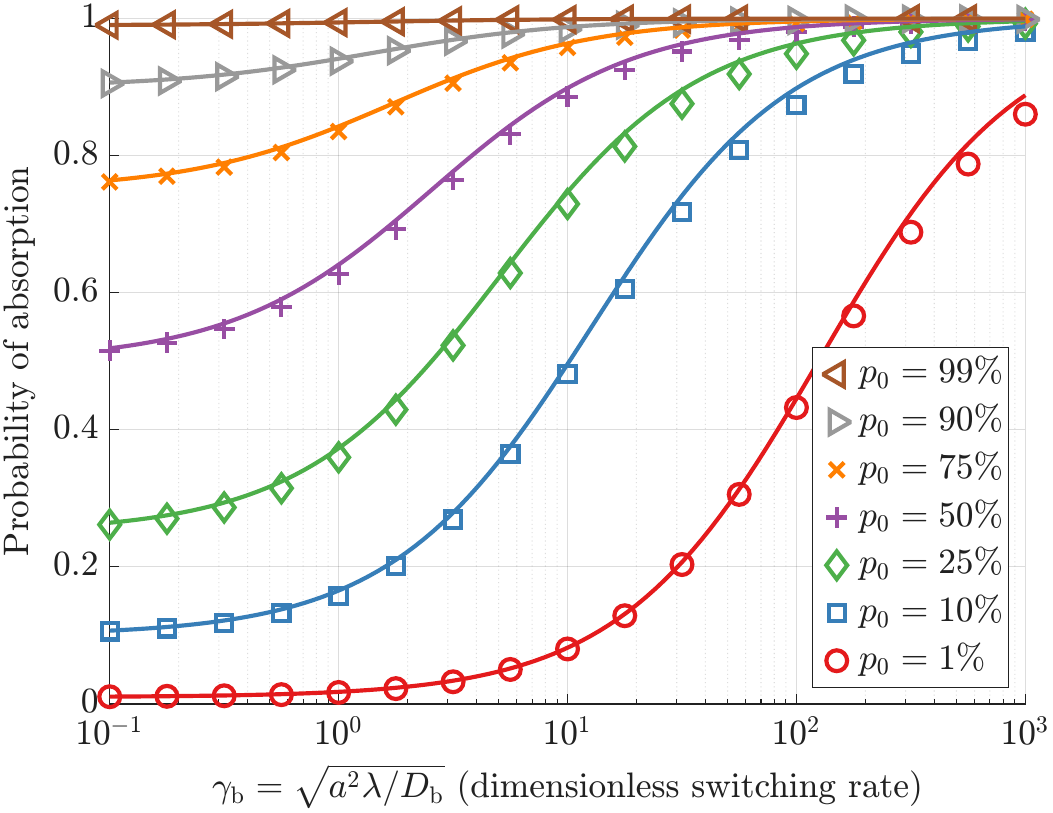}
\caption{Comparison of $\mathcal{P}_{\text{approx}}$ in \eqref{eq:Prho0} (curves) to stochastic simulations (markers).}
\label{fig:diskgated}
\end{figure}

Figure~\ref{fig:diskgated} compares our approximation \eqref{eq:Prho0} (solid curves) to stochastic simulations (markers) for a range of values of the open probability $p_0$ and the dimensionless switching rate $\gb=\sqrt{\lambda a^2/\Db}$. This plot shows good agreement, with the relative error never exceeding 5\% and the absolute error never exceeding 0.015. This figure illustrates that the splitting probability is merely the open probability if switching is slow,
\begin{align*}
    \lim_{\gb\to0}\mathcal{P}_{\text{approx}}=\lim_{\gb\to0}\mathcal{P}=p_0,
\end{align*}
whereas the splitting probability is unity if switching is fast,
\begin{align*}
    \lim_{\gb\to\infty}\mathcal{P}_{\text{approx}}=\lim_{\gb\to\infty}\mathcal{P}=1.
\end{align*}

\begin{figure*}[t]
\centering
\includegraphics[width=1\linewidth]{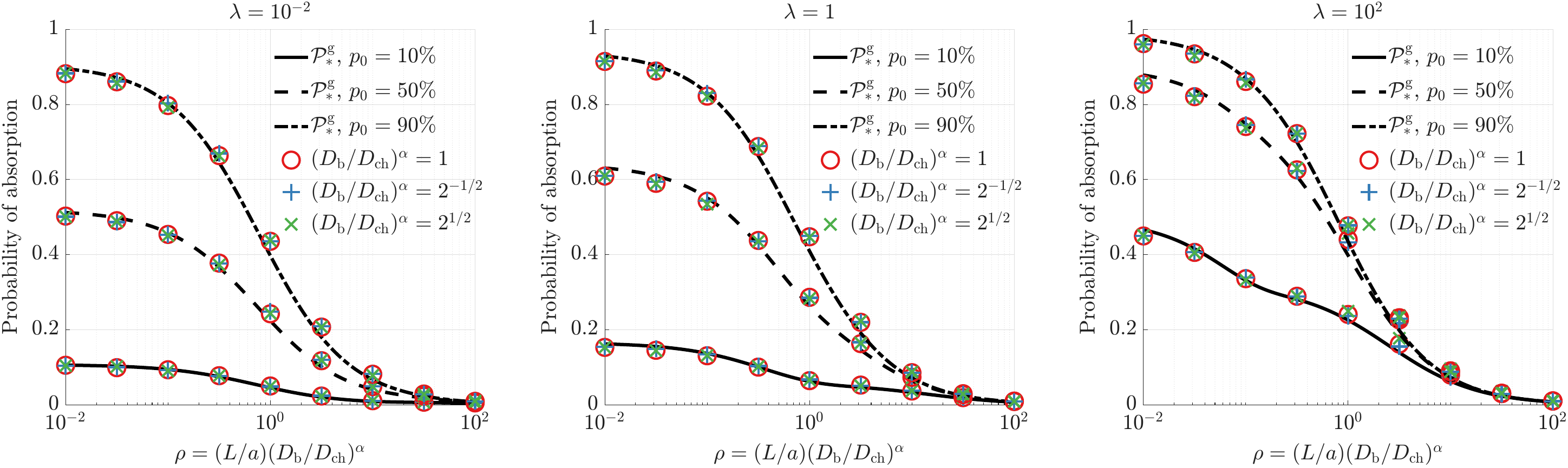}
\caption{Comparison of $\mathcal{P}_{\text{approx}}$ in \eqref{eq:Pmain} (black curves) to stochastic simulations (markers). The red circle markers are for $\alpha=0$. The blue $+$ markers and green $\times$ markers are both for $\alpha=1/2$.}
\label{fig:gated}
\end{figure*}

Having established that $\mathcal{P}_{\text{approx}}$ in \eqref{eq:Pmain} is (a) exact for $\rho\gg1$ (shown analytically in section~\ref{sec:tube}) and (b) accurate for $\rho=0$ (in Figure~\ref{fig:diskgated}), we now consider intermediate values of $\rho$. Figure~\ref{fig:gated} plots $\mathcal{P}_{\text{approx}}$ in \eqref{eq:Pmain} as a function of $\rho=(L/a)(\Db/\Dc)^\alpha$ for different values of $p_0$ (black curves) against stochastic simulations (markers) for different choices of the ratio $\Db/\Dc$ and the multiplicative noise parameter $\alpha$. The three panels are for different values of the switching rate $\lambda$. Figure~\ref{fig:diskgated} shows good agreement between the theoretical approximation and simulations.

\section{\label{sec:predictions}Predictions of flux formula}

We now explore the predictions of the approximate flux formula derived in section~\ref{sec:tube}. For simplicity, suppose that the tube is a cylinder so that the dimensionless geometric factor appearing in our analysis is
\begin{align*}
    G(\Gamma)
    =\frac{2\pi C_0(\Gamma)}{|\Gamma|}
    =\frac{4}{\pi}.
\end{align*}
We use the letter $J$ to denote a flux, where the superscript ``open'' denotes the case that the gate is always open (no superscript denotes the case that the entrance to the tube is stochastically gated). Furthermore, we use the subscript ``exact'' for the exact flux (which is known only in some circumstances) and other subscripts for various approximations (specified below). 

The analysis in section~\ref{sec:tube} predicts that the exact ``always open'' flux $J_{\text{exact}}^{\text{open}}$ is well-approximated by 
\begin{align}\label{eq:fng}
    J_{\text{exact}}^{\text{open}}
    \approx J_*^{\text{open}}
    &:=\frac{4a\Db c_\infty}{1+\frac{4}{\pi}\rho},
\end{align}
and that the exact gated flux $J_{\text{exact}}$ is well-approximated by
\begin{align}\label{eq:fg}
    J_{\text{exact}}
    \approx J_*
    &:=\frac{4a\Db c_\infty}{\frac{\frac{4}{\pi}\frac{1}{p_0}+{\gb}}{\frac{4}{\pi}+{\gb}}+\frac{4}{\pi}\rho(1+\frac{p_1}{p_0}\frac{\tanh(\gamma)}{\gamma})},
\end{align}
where $\rho$ compares the tube aspect ratio to the change in diffusivity at the tube entrance,
\begin{align}\label{eq:rhodefpredict}
    \rho
    =\frac{L}{a}\Big(\frac{\Db}{\Dc}\Big)^\alpha\ge0,
\end{align}
$p_0$ and $p_1=1-p_0$ denote the respective fractions of time that the gate is open and closed, and $\gamma$ and ${\gb}$ compare the switching rate $\lambda:=\lambda_0+\lambda_1$ to the timescales of diffusion in the tube and the bulk,
\begin{align}\label{eq:gammas}
    \gamma=\sqrt{\lambda L^2/\Dc},\quad
    {\gb}=\sqrt{\lambda a^2/\Db}. 
\end{align}
The accuracy of the approximations in \eqref{eq:fng}-\eqref{eq:fg} is supported by the simulations in section~\ref{sec:simulations}. We now investigate the predictions of the flux formulas in \eqref{eq:fng}-\eqref{eq:fg}.

\subsection{Slow and fast switching}

It follows immediately from \eqref{eq:fng}-\eqref{eq:fg} that if the switching rate is much slower than the diffusion timescales, then the gated flux is merely the ``always open'' flux multiplied by the fraction of time that the gate is open,
\begin{align}\label{eq:slow}
    \lim_{\gamma,{\gb}\to0}J_*=p_0J_*^{\text{open}}.
\end{align}
More interestingly, \eqref{eq:fng}-\eqref{eq:fg} imply that the gated flux is identical to the always open flux if the switching is much faster than the diffusion timescales,
\begin{align}\label{eq:fast}
    \lim_{\gamma,{\gb}\to\infty}J_*=J_*^{\text{open}}.
\end{align}
We emphasize that \eqref{eq:slow} and \eqref{eq:fast} each hold for any value of $\rho\ge0$. We further note that \eqref{eq:fast} contrasts the predictions of Ref.~\citenum{berezhkovskii2016diffusive} (see section~\ref{sec:discussion} below).

\subsection{Large $\rho=(L/a)(\Db/\Dc)^\alpha$}

In the large $\rho$ limit, the fluxes in \eqref{eq:fng}-\eqref{eq:fg} have the following exact asymptotics,
\begin{align}
    J_{\text{exact}}^{\text{open}}
    &\sim J_*^{\text{open}}
    \sim\frac{\pi a\Db c_\infty}{\rho}
    \sim\pi a^2  \frac{\Db^{1-\alpha}\Dc^\alpha c_\infty}{L},\quad\text{as }\rho\to\infty,\label{eq:largerhofng}\\
    J_{\text{exact}}
    &\sim J_*
    \sim\frac{J_*^{\text{open}}}{1+\frac{p_1}{p_0}\frac{\tanh(\gamma)}{\gamma}},\quad\,\,\quad\text{as }\rho\to\infty,\label{eq:largerhofg}
\end{align}
where $g\sim h$ denotes $\lim g/h=1$. 
We make four comments about \eqref{eq:largerhofng}-\eqref{eq:largerhofg}. 

First, \eqref{eq:largerhofng}-\eqref{eq:largerhofg} means that the flux formulas in \eqref{eq:fg}-\eqref{eq:fng} are exact in the $\rho\gg1$ limit (see Appendix~\ref{sec:largerho}).

Second, \eqref{eq:largerhofg} generalizes the result in \eqref{eq:thesis} of Refs.~\citenum{lawleythesis} and \citenum{lawley15sima} to the case that the bulk diffusivity $\Db$ and the tube diffusivity $\Dc$ may differ.

Third, the fluxes in \eqref{eq:largerhofng}-\eqref{eq:largerhofg} depend strongly on the multiplicative noise interpretation parameter $\alpha\in[0,1]$ if $\Db\neq\Dc$. 

Fourth, the large parameter in \eqref{eq:largerhofng}-\eqref{eq:largerhofg} is $\rho$ in \eqref{eq:rhodefpredict}, which compares the tube aspect ratio $L/a$ to the change in diffusivity at the tube entrance $\Db/\Dc$, where the sensitivity to the change is diffusivity is weighted by $\alpha\in[0,1]$. Hence, the asymptotics in \eqref{eq:largerhofng}-\eqref{eq:largerhofg} may hold even for a short tube (i.e.\ \eqref{eq:largerhofng}-\eqref{eq:largerhofg} holds even for $L/a\ll1$ if $(\Db/\Dc)^\alpha$ is sufficiently large). This point can be understood by noting that \eqref{eq:largerhofng}-\eqref{eq:largerhofg} correspond to the parameter regime that a diffusing particle which enters the tube is unlikely to traverse the tube. This can happen because the tube is long ($L/a\gg1$) and/or the diffusivity is much faster in the bulk than the tube and $\alpha\neq0$ ($(\Db/\Dc)^\alpha\gg1$), since diffusing particles are biased toward regions of large diffusivity if $\alpha>0$.

\subsection{Small $\rho=(L/a)(\Db/\Dc)^\alpha$}

In the small $\rho$ limit, the fluxes in \eqref{eq:fng}-\eqref{eq:fg} reduce to
\begin{align}
    J_*^{\text{open}}
    &\to J_{\text{disk}}^{\text{open}}:=4a\Db c_\infty\quad\quad\quad\;\quad\text{as }\rho\to0,\label{eq:smallrhofng}\\
    J_*
    &\to J_{\text{disk}}:=\bigg(\frac{\frac{4}{\pi}+{\gb}}{\frac{4}{\pi}\frac{1}{p_0}+{\gb}}\bigg)J_{\text{disk}}^{\text{open}}\quad\text{as }\rho\to0.\label{eq:smallrhofg}
\end{align}
We make three comments about \eqref{eq:smallrhofng}-\eqref{eq:smallrhofg}.

First, the non-gated flux formula $J_{\text{disk}}^{\text{open}}$ in \eqref{eq:smallrhofng} is exact in the small $\rho$ limit (i.e.\ $\lim_{\rho\to0}J_{\text{exact}}^{\text{open}}=\lim_{\rho\to0}J_{*}^{\text{open}}=J_{\text{disk}}^{\text{open}}$). Indeed, the non-gated flux $J_{\text{exact}}^{\text{open}}$ in the small $\rho$ limit is merely the diffusive flux to a perfectly absorbing disk, which is well-known\cite{berg1977} to be $4a \Db c_\infty$. 

Second, the gated flux formula $J_{\text{disk}}$ in \eqref{eq:smallrhofg} is not exact in the sense that $\lim_{\rho\to0}J_{\text{exact}}\neq J_{\text{disk}}$. This is not surprising, since $\lim_{\rho\to0}J_{\text{exact}}$ is the flux to a gated disk, which is notoriously difficult to estimate\cite{szabo1982}. Nevertheless, the simulations in section~\ref{sec:simulations} (see Figure~\ref{fig:diskgated}) show that the formula $J_{\text{disk}}$ in \eqref{eq:smallrhofg} is a very accurate approximation of the exact flux,
\begin{align*}
    \lim_{\rho\to0}J_{\text{exact}}\approx J_{\text{disk}}.
\end{align*}

Third, the gated flux formula $J_{\text{disk}}$ in \eqref{eq:smallrhofg} still depends on the switching rate $\lambda$ via the parameter ${\gb}=\sqrt{\lambda a^2/\Db}$. 
Indeed, $J_{\text{disk}}$ predicts that the gated flux is equal to $p_0 J_{\text{disk}}^{\text{open}}$ only in the slow switching limit,
\begin{align}\label{eq:f01}
    \lim_{{\gb}\to0}J_{\text{disk}}=p_0J_{\text{disk}}^{\text{open}}.
\end{align}
Furthermore, $J_{\text{disk}}$ predicts that the gated flux is always strictly larger than the estimate $p_0 J_{\text{disk}}^{\text{open}}$,
\begin{align}\label{eq:f02}
    J_{\text{disk}}>p_0J_{\text{disk}}^{\text{open}}\quad\text{for any }{\gb}>0,\,p_0\in(0,1).
\end{align}
In fact, \eqref{eq:smallrhofg} predicts that the gated flux is identical to the non-gated flux $J_{\text{disk}}^{\text{open}}$ in the fast switching limit, 
\begin{align}\label{eq:f03}
    \lim_{{\gb}\to\infty}J_{\text{disk}}=J_{\text{disk}}^{\text{open}}.
\end{align}
It is straightforward to prove that the predictions in \eqref{eq:f01}-\eqref{eq:f03} are correct in the sense that \eqref{eq:f01}-\eqref{eq:f03} each hold with $J_{\text{disk}}$ replaced by the exact gated flux for a tube of zero depth. That is, \eqref{eq:f01}-\eqref{eq:f03} hold with $J_{\text{disk}}$ replaced by $\lim_{\rho\to0}J_{\text{exact}}$. 

\subsection{\label{sec:opposite3D} Fast and slow switching on opposite sides of the gate}

As in section~\ref{sec:opposite1D} for the one-dimensional, exactly solvable example, the difference in diffusivities ($\Db$ and $\Dc$) and relevant lengthscales ($a$ and $L$) in the bulk versus the tube mean that the effective switching rates can be very different on opposite sides of the gate (i.e.\ it is possible to have $\gamma\gg\gb$ or $\gamma\ll\gb$). This can lead to some non-intuitive results for the approximation $\mathcal{P}_{\text{approx}}$ in \eqref{eq:Pmain} and thus the flux $J_*$.

\begin{figure}[t]
\centering
\includegraphics[width=1\linewidth]{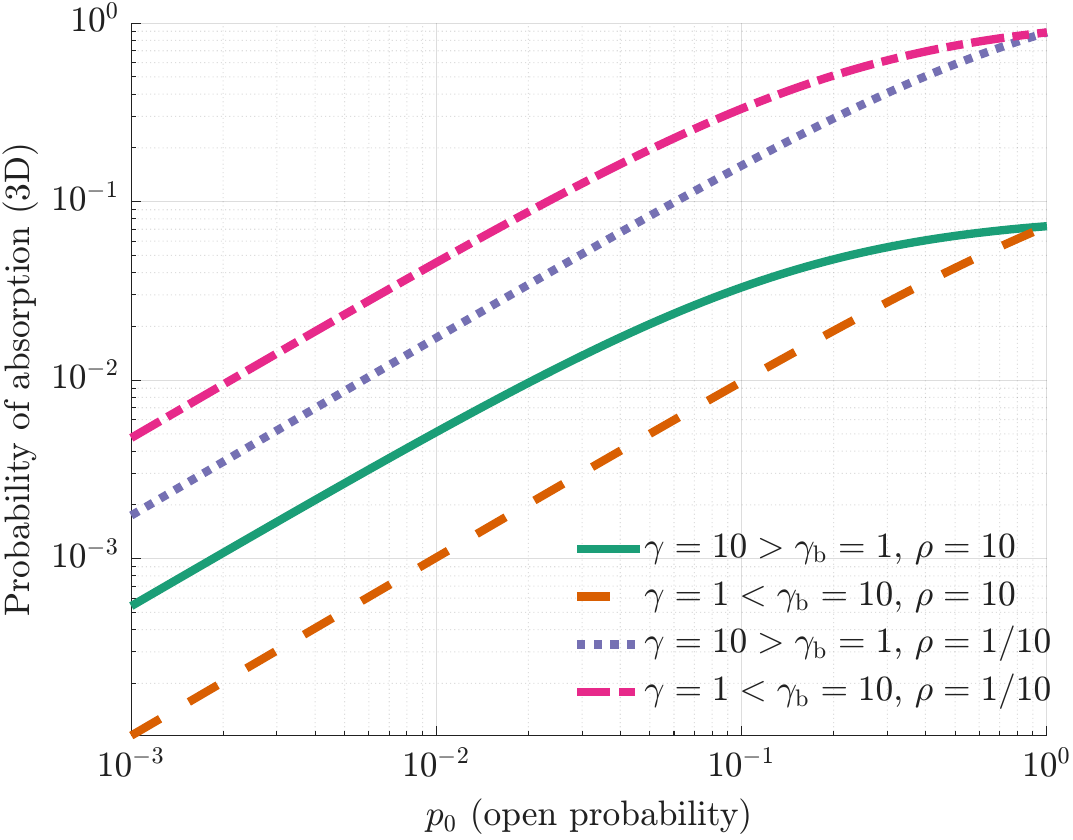}
\caption{Plot of $\mathcal{P}_{\text{approx}}$ in \eqref{eq:Pmain} as a function of the open probability $p_0$ in different parameter regimes. See section~\ref{sec:opposite3D} for details.}
\label{fig:fastslow3D}
\end{figure}

This is illustrated in Figure~\ref{fig:fastslow3D}, which plots $\mathcal{P}_{\text{approx}}$ in \eqref{eq:Pmain} for the case of a cylindrical channel (so that $G(\Gamma)=4/\pi$). Note that Figure~\ref{fig:fastslow3D} is nearly identical to Figure~\ref{fig:fastslow1D}, where Figure~\ref{fig:fastslow1D} concerns the exactly solvable one-dimensional (1D) example considered in section~\ref{sec:firstpedagogical}. As in section~\ref{sec:opposite1D}, the inequalities in \eqref{eq:largerho1D}-\eqref{eq:smallrho1D} hold with $P(0^-)$ replaced by $\mathcal{P}_{\text{approx}}$, and the reasoning behind \eqref{eq:largerho1D}-\eqref{eq:smallrho1D} still holds for $\mathcal{P}_{\text{approx}}$. Furthermore, the various parameter regimes in \eqref{eq:regime1}-\eqref{eq:regime4} can be equivalently obtained for this three-dimensional model. 

\section{\label{sec:discussion}Discussion}

We conclude by comparing our flux estimate $J_*$ to prior estimates of the diffusive flux into a stochastically gated tube\cite{lawleythesis, lawley15sima, berezhkovskii2016diffusive}. To facilitate comparison, we assume that the tube is a cylinder of radius $a$ and length $L$.

To the best of our knowledge, the flux into a stochastically gated tube was first studied in Refs.~\citenum{lawleythesis,lawley15sima}. In Refs.~\citenum{lawleythesis,lawley15sima}, the entrance to the tube at $x=0$ was assumed to open and close according to \eqref{eq:markovsimple} and particles have diffusivity $\Dc$. Assuming that the tube is narrow (i.e.\ $a/L\ll1$), the following flux estimate was derived by analyzing a one-dimensional stochastic partial differential equation,
\begin{align}\label{eq:thesis9}
        J_{2014}
    :=\frac{4a\Dc c_\infty}{(1+\frac{p_1}{p_0}\frac{\tanh(\gamma)}{\gamma})\frac{4}{\pi}(L/a)},
\end{align}
where $\gamma=\sqrt{\lambda L^2/\Dc}$ compares the switching rate to the timescale of diffusion in the tube, and $c_\infty$ is the ambient particle concentration in the bulk reservoir (\eqref{eq:thesis9} results from identifying the constant $b$ in Refs.~\citenum{lawleythesis, lawley15sima} with the product of the tube cross-sectional area $\pi a^2$ and the bulk concentration $c_\infty$). 

A very interesting work\cite{berezhkovskii2016diffusive} sought to extend the flux estimate in \eqref{eq:thesis9} to be valid if 
\begin{enumerate}[label=(\roman*), start=1]
    \item the tube is not necessarily narrow (i.e.\ $L\not\gg a$),
    \item the diffusivity differs in the tube versus the bulk.
\end{enumerate}
By devising an elegant formalism and analyzing a resulting reaction-diffusion problem, Ref.~\citenum{berezhkovskii2016diffusive} derived the following flux formula (see Equation~(4.1) in Ref.~\citenum{berezhkovskii2016diffusive}), 
\begin{align}\label{eq:J16}
    J_{2016}
    &:=\frac{4a\Db c_\infty}{1+\frac{p_1}{p_0}+(1+\frac{p_1}{p_0}\frac{\tanh(\gamma)}{\gamma})\frac{4}{\pi}(L/a)(\Db/\Dc)},
\end{align}
where $\Db$ and $\Dc$ denote the respective bulk and tube diffusivities and $\gamma=\sqrt{\lambda L^2/\Dc}$. 
In this paper, we adopted an alternative approach and derived the following flux estimate,
\begin{align}\label{eq:Jstar}
    J_*
    &:=\frac{4a\Db c_\infty}{1+\frac{p_1}{p_0}\frac{1}{1+{\gb}\pi/4}+(1+\frac{p_1}{p_0}\frac{\tanh(\gamma)}{\gamma})\frac{4}{\pi}(L/a)(\Db/\Dc)^\alpha},    
\end{align}
where $\alpha\in[0,1]$ specifies the form of the multiplicative noise inherent to any space-dependent diffusivity\cite{mannella2012ito} and ${\gb}=\sqrt{\lambda a^2/\Db}$ compares the switching rate $\lambda$ to the timescale of diffusion in the bulk $a^2/\Db$.

\subsection{Narrow tube and constant diffusivity}

The flux estimate $J_{2014}$ was derived assuming that the tube is narrow ($a/L\ll1$) and the particle diffusivity is constant ($\Dc=\Db$). The newer flux estimates $J_{2016}$ and $J_{*}$ in \eqref{eq:J16}-\eqref{eq:Jstar} reduce to $J_{2014}$ under these assumptions. Indeed, it is straightforward to check that
\begin{align*}
    J_{2014}
    \sim J_{2016}
    \sim J_*
    \quad\text{as }a/L\to0\;\text{ if $\Db=\Dc$},
\end{align*}
where $f\sim g$ denotes $f/g\to1$.

\subsection{Short tube (extension (i))}

We now consider extension (i) by removing the assumption that $a/L\ll1$. Specifically, in the case of a short tube, $J_{2016}$ in \eqref{eq:J16} reduces to
\begin{align}\label{eq:J16wide}
    \lim_{L/a\to0}J_{2016}
    =p_0J_{\text{disk}}^{\text{open}}
    =p_0(4a\Db c_\infty),
\end{align}
which contrasts $J_*$ in \eqref{eq:Jstar},
\begin{align}\label{eq:Jstarwide}
    \lim_{L/a\to0}J_*
    =\Big(p_0+\frac{p_1p_0{\gb}}{4/\pi+p_0{\gb}}\Big)(4a\Db c_\infty).
\end{align}
Notice that \eqref{eq:J16wide} depends on the stochastic gate only via the open probability $p_0$. In contrast, \eqref{eq:Jstarwide} depends on both $p_0$ and the dimensionless parameter ${\gb}=\sqrt{\lambda a^2/\Db}$, which compares the switching rate $\lambda$ to the timescale of diffusion in the bulk. The estimates \eqref{eq:J16wide} in \eqref{eq:Jstarwide} are identical in the slow switching limit (${\gb}\to0$) and in the always open limit ($p_0=1-p_1\to1$). The discrepancy between \eqref{eq:J16wide} and \eqref{eq:Jstarwide} is most pronounced if
\begin{align*}
    p_0\ll1\ll p_0{\gb}.
\end{align*}
in which case the estimate \eqref{eq:Jstarwide} predicts that the gated flux is identical to the flux to a perfectly absorbing disk (since $p_0{\gb}\gg1$, i.e.\ switching is fast), which is much greater than \eqref{eq:J16wide} if $p_0\ll1$.

The accuracy of \eqref{eq:Jstarwide} is supported by Figure~\ref{fig:diskgated} since the problem is equivalent to calculating the diffusive flux to a stochastically gated disk if $L/a\to0$. Furthermore, the diffusive flux to a stochastically gated disk was studied by Szabo et al.~\cite{szabo1982}, who predicted that (1) this gated flux limits to $p_0(4a\Db c_\infty)$ for slow switching (${\gb}\to0$) and (2) this gated flux limits to $4a\Db c_\infty$ for fast switching (${\gb}\to\infty$), which both agree with \eqref{eq:Jstarwide}.

\subsection{Different bulk and tube diffusivities (extension (ii))}

We now consider extension (ii) by supposing that $\Db\neq\Dc$. If the tube is narrow, then 
\begin{align}\label{eq:noalpha}
    J_{2016}
    \sim\frac{\Dc}{\Db}J_{2014}\quad\text{as }a/L\to0,
\end{align}
which contrasts
\begin{align}\label{eq:yesalpha}
    J_{*}
    \sim\Big(\frac{\Dc}{\Db}\Big)^\alpha J_{2014}\quad\text{as }a/L\to0.
\end{align}
Hence, \eqref{eq:yesalpha} predicts that the flux can depend strongly on the interpretation of the multiplicative noise inherent to any space-dependent diffusivity. Furthermore, \eqref{eq:yesalpha} predicts that the narrow tube regime of $J_{2016}$ (given in \eqref{eq:noalpha}) is valid if $\alpha=1$, which is called the kinetic or isothermal interpretation of multiplicative noise\cite{hanggi1982stochastic}. 

However, even if $\alpha=1$, the flux estimate $J_{*}$ in \eqref{eq:Jstar} generally differs from $J_{2016}$ in \eqref{eq:J16}. To illustrate, suppose 
\begin{align}\label{eq:kinetic}
    \alpha=1\quad\text{and}\quad
    (L/a)(\Db/\Dc)\ll1,
\end{align}
in which case the problem is again equivalent to calculating the flux to a stochastically gated disk. To see this equivalence, note that if \eqref{eq:kinetic} holds, then any particle which reaches the gate when it is open will very likely reach the end of the tube because the tube is short ($L/a\ll1$) and/or the small bulk versus tube diffusivity ($\Db/\Dc\ll1$) and the kinetic interpretation of multiplicative noise ($\alpha=1$) will push the particle into the tube and strongly inhibit its return to the bulk. 
In this case,
\begin{align}\label{eq:J16wide2}
    \lim_{(L/a)(\Db/\Dc)\to0}J_{2016}
    =p_0J_{\text{disk}}^{\text{open}}
    =p_0(4a\Db c_\infty),
\end{align}
which contrasts $J_*$,
\begin{align}\label{eq:Jstarwide2}
    \lim_{(L/a)(\Db/\Dc)\to0}J_*
    =\Big(p_0+\frac{p_1p_0{\gb}}{4/\pi+p_0{\gb}}\Big)(4a\Db c_\infty).
\end{align}
The discrepancy between \eqref{eq:J16wide2} and \eqref{eq:Jstarwide2} mirrors the discrepancy between \eqref{eq:J16wide} and \eqref{eq:Jstarwide}, and the accuracy of \eqref{eq:Jstarwide2} is again supported by Figure~\ref{fig:diskgated} and the predictions of Szabo et al.~\cite{szabo1982}

\subsection{Fast switching}

Finally, another salient discrepancy between $J_{2016}$ in \eqref{eq:J16} versus $J_{2014}$ in \eqref{eq:thesis9} and $J_*$ in \eqref{eq:Jstar} is in the fast switching limit. In particular, if the switching rate $\lambda$ is much faster than the timescales of diffusion in the problem, then both $J_{2014}$ and $J_*$ predict that the flux is the same as the flux for the case that the gate is always open,
\begin{align*}
    \lim_{\gamma\to\infty}J_{2014}
    &=\lim_{p_0\to1}J_{2014},\\
    \lim_{\gamma,\gb\to\infty}J_{*}
    &=\lim_{p_0\to1}J_{*},
\end{align*}
which contrasts $J_{2016}$ (see equations (4.8)-(4.9) in Ref.~\citenum{berezhkovskii2016diffusive}),
\begin{align*}
    \lim_{\gamma\to\infty}J_{2016}
    <\lim_{p_0\to1}J_{2016}.
\end{align*}


\section*{Acknowledgments}

The author gratefully acknowledges support from the National Science Foundation (NSF DMS-2325258).

\section*{Author Declarations}

The author has no conflicts to disclose.

\section*{Data availability}

The data that support the findings of this study are available from the corresponding author upon reasonable request.

\appendix

\section{\label{sec:largerho}Approximations are exact for $\rho\gg1$}

Owing to the exact relation in \eqref{eq:exactswitch}, we need only that \eqref{eq:P0vanish} holds in order to establish that the approximation $\mathcal{P}_{\text{approx}}$ is exact as $\rho\to\infty$ (i.e.\ \eqref{eq:P0vanish} implies \eqref{eq:exactasymptotic}). The result \eqref{eq:P0vanish} is intuitive, as it merely says that the probability that a particle starting at the open gate is unlikely to reach the end of the tube if the tube is long and/or the diffusivity is much faster in the bulk than the tube and $\alpha>0$.

To make this argument precise, observe that the strong Markov property implies that
\begin{align*}
    P_0(0,y,z)
    &\le P_{\text{1d}}
    +\sup_{(y',z')\in\R^2}P_0(-a,y',z'),
\end{align*}
where $P_{\text{1d}}$ is the one-dimensional splitting probability in \eqref{eq:Pgatedsimple},
\begin{align*}
    P_{\text{1d}}
    =\Big[1+\frac{p_1}{p_0}\frac{\tanh ({\gb})}{{\gb}}+\rho \Big(1+\frac{p_1}{p_0}\frac{\tanh (\gamma)}{\gamma}\Big)\Big]^{-1}.
\end{align*}
The strong Markov property further implies that
\begin{align*}
    \sup_{(y,z)\in\R^2}P_0(-a,y,z)
    \le\sup_{(y,z)\in\R^2}h(-a,y,z;a\Gamma)\sup_{(y,z)\in\R^2}P_0(0,y,z),
\end{align*}
where $h(x,y,z;a\Gamma)$ is the probability that $\X$ hits $a\Gamma$ starting from $(x,y,z)$. 
Therefore,
\begin{align*}
    \sup_{(y,z)\in\R^2}P_0(0,y,z)
    \le\frac{P_{\text{1d}}}{1-\sup_{(y,z)\in\R^2}h(-a,y,z;a\Gamma)}.
\end{align*}
Furthermore, rescaling space by $a$ yields
\begin{align*}
    \sup_{(y,z)\in\R^2}h(a,y,z;a\Gamma)=\sup_{(y,z)\in\R^2}h(1,y,z;\Gamma)
    \neq1,
\end{align*}
which is independent of $a$, $L$, $\Db$, $\Dc$, and $\alpha$ (and thus independent of $\rho$). Hence, $\sup_{(y,z)\in\R^2}h(a,y,z;a\Gamma)$ is bounded away from 1 for all $\rho>0$, and therefore
\begin{align*}
    \lim_{\rho\to\infty}\sup_{(y,z)\in\R^2}P_0(0,y,z)=0,
\end{align*}
since $\lim_{\rho\to\infty}P_{\text{1d}}=0$. Since $\overline{P}_0(0)\le\sup_{(y,z)\in\R^2}P_0(0,y,z)$, we conclude that \eqref{eq:P0vanish} holds.

\section{\label{sec:derivationofkey}Derivation of \eqref{eq:Qintapprox}}

We now derive \eqref{eq:Qintapprox}, which is equivalent to
\begin{align}\label{eq:Qadd}
    \int_{x<0} Q\,\dd \x
    &\approx\frac{2\pi a C_0(\Gamma)+{\gb}a|\Gamma|}{\lambda/\Db}\overline{Q}(0^-).
\end{align}
The approximation in \eqref{eq:keyusual} ignores variation in the channel cross-section and is equivalent to
\begin{align}\label{eq:keyusualQ}
    Q(0^-,y,z)\approx\overline{Q}(0^-)\quad\text{for all }(y,z)\in a\Gamma.
\end{align}
For ${\gb}=\sqrt{a\lambda/\Db}\gg1$, the solution $Q$ of
\begin{align}\label{eq:QPDEcoordinates}
    \Delta Q=
    (\partial_{xx}+\partial_{yy}+\partial_{zz})Q=(\lambda/\Db) Q,
\end{align}
vanishes in the bulk outside of a small boundary layer of the entrance to the tube. Furthermore, the approximation \eqref{eq:keyusualQ} means that the $y$ and $z$ derivatives of $Q$ vanish inside this boundary layer, and thus we approximate the solution to \eqref{eq:QPDEcoordinates} by the solution to $\partial_{xx}Q=(\lambda/\Db) Q$, which is $Q(x,y,z)=Q(0,y,z)e^{-x\sqrt{\lambda/\Db}}$. Therefore, we arrive at the following approximation for large ${\gb}$,
\begin{align}\label{eq:Qlarge}
    \int_{x<0}Q\,\dd x
    \approx a^2|\Gamma|\overline{Q}(0^-)\int_0^\infty e^{-x\sqrt{\lambda/\Db}}\,\dd x
    =\frac{{\gb}a|\Gamma|}{\lambda/\Db}\overline{Q}(0^-).
\end{align}

To obtain an approximation for ${\gb}=\sqrt{a\lambda/\Db}\ll1$, we integrate \eqref{eq:QPDEcoordinates} over the entire bulk region and use the divergence theorem to obtain
\begin{align}\label{eq:Qsmall}
    \frac{\lambda}{\Db}\int_{x<0}Q\,\dd \x
    =\int_{a\Gamma}\partial_x Q(0^-,y,z)\,\dd y\,\dd z
    \approx 2\pi aC_0(\Gamma)\overline{Q}(0^-),
\end{align}
where the approximation comes from noting that the solution $Q$ to \eqref{eq:QPDEcoordinates} with boundary condition \eqref{eq:keyusualQ} approaches the solution to Laplace's equation as ${\gb}\to0$ with boundary condition \eqref{eq:keyusualQ}.

Simply adding the large ${\gb}$ approximation in \eqref{eq:Qlarge} and the small ${\gb}$ approximation in \eqref{eq:Qsmall} yields the approximation in \eqref{eq:Qadd} (which is equivalent to \eqref{eq:Qintapprox}), which reduces to \eqref{eq:Qlarge} as ${\gb}\to\infty$ and \eqref{eq:Qsmall} as ${\gb}\to0$.

\section{\label{sec:algorithm}Stochastic simulation algorithm}

We now describe the stochastic simulation algorithms used in section~\ref{sec:simulations}.

\subsection{\label{sec:algorithmdisk}Gated disk}

We start in the case that the cylindrical tube has zero length ($L=0$). In this case, we need to compute the probability that a particle diffusing in the bulk gets absorbed at a stochastically gated disk, given that it reaches the disk from far away in the bulk. The following stochastic simulation algorithm is used to compute this probability and produce Figure~\ref{fig:diskgated}. Each marker in Figure~\ref{fig:diskgated} is computed from $5\times10^6$ independent simulations.

\textbf{Step 0.} Initialize the gate to be open with probability $p_0$ and closed with probability $p_1=1-p_0$. If the gate is initially open, then the particle is absorbed and the simulation ends. If the gate is initially closed, then initialize the particle to be at the following position on the disk,
\begin{align}\label{eq:initial}
    \X(0)
    =(X(0),Y(0),Z(0))
    =(0,0,a\sqrt{1-U^2}),
\end{align}
where $U$ is uniformly distributed on $(0,1)$. The initial position \eqref{eq:initial} generates statistically exact samples from \eqref{eq:hittingdensity}. Proceed to Step 1. 

\textbf{Step 1.} Given the current position of the particle $\X_0=(0,Y_0,Z_0)\in a\Gamma$ and that the gate is closed, sample the waiting time until the gate opens, which is given exactly by
\begin{align*}
    \tau
    =E/\lambda_1,
\end{align*}
where $E$ is a unit mean exponential random variable. Until time $\tau$ elapses, the particle diffuses in three dimensions with a reflecting boundary condition at $x=0$, and therefore its position when the gate opens is exactly given by
\begin{align}
    X
    &=-|0+\sqrt{2\Db\tau}\mathcal{N}_x|,\label{eq:X}\\
    Y
    &=Y_0+\sqrt{2\Db\tau}\mathcal{N}_y,\label{eq:Y}\\
    Z
    &=Z_0+\sqrt{2\Db\tau}\mathcal{N}_z,\label{eq:Z}
\end{align}
where $\mathcal{N}_x,\mathcal{N}_y,\mathcal{N}_z,$ are independent standard normal random variables.

\textbf{Step 2:} Starting from position $\X=(X,Y,Z)$ in \eqref{eq:X}-\eqref{eq:Z}, follow the algorithm in Ref.~\citenum{bernoff2018boundary} to simulate the path of the particle until it either (a) reaches a distance $R_\infty=10^{10}\gg1$ from the disk or (b) hits $a\Gamma$ at $x=0$ (i.e.\ the particle hits the disk). If (a) occurs, then we assume that the particle will never reach the disk and the simulation ends. If (b) occurs, then set the gate to be closed with probability 
\begin{align}\label{eq:closedsimulation}
    p_1(1-e^{-(\lambda_0+\lambda_1)t}),
\end{align}
where $t$ is the time elapsed since the start of Step 2. If the gate is open, then the particle is absorbed and the simulation ends. If the gate is closed, then return to Step 1.

\subsection{Gated tube}

The following stochastic simulation algorithm is used to produce Figure~\ref{fig:gated}. Each marker Figure~\ref{fig:gated} is computed from $8\times10^4$ independent simulations.

\textbf{Step 1.} Follow the algorithm in section~\ref{sec:algorithmdisk} above until either the particle (a) reaches a distance $R_\infty\gg1$ from the tube opening or (b) hits the entrance to the tube when the gate is open. If (a) occurs, then we assume that the particle will never reach the entrance to the tube and the simulation ends.  If (b) occurs, then proceed to Step 2. 

\textbf{Step 2.} Introduce a small simulation parameter $\eta>0$ and set the new position of the particle to be
\begin{align*}
    X=X_0\pm\eta,\quad Y=Y_0,\quad Z=Z_0,
\end{align*}
where
\begin{align*}
    \P(X=X_0+\eta)
    =P_0(0)=1-\P(X=X_0-\eta),
\end{align*}
and $P_0(0)$ is calculated from \eqref{eq:Pgatedsimple} with $L=a=\eta$. In our simulations, we take $\eta=\min\{0.1,L/2\}$.

\textbf{Step 3.} If $X<0$, then the particle is in the bulk, and we return to Step 1. If $X>0$, then the particle is in the tube, and we proceed to Step 4.

\textbf{Step 4.} Sample the first time $t$ that the $X$ coordinate of the particle escapes $(0,L)$ and whether it escapes at $x=0$ or $x=L$, which can be done as in Appendix~A of Ref.~\citenum{richardson2026yet}. If it escapes at $x=L$, then the particle is absorbed and the simulation ends. If it hits $x=0$, then set the gate to be closed with probability \eqref{eq:closedsimulation} and sample the position of the $Y$ and $Z$ coordinates of the particle, which can be done as in Appendix~A of Ref.~\citenum{richardson2026yet}. If the gate is open, return to Step 2. If the gate is closed, then update $X$ to be $\eta>0$ and return to the start of Step 4. 


\bibliography{library}
\bibliographystyle{unsrt}

\end{document}